\documentclass[reprint,onecolumn,nofootinbib,notitlepage]{revtex4-2}
\usepackage[utf8]{inputenc}
\usepackage[top=2cm,bottom=3.5cm,left=2cm,right=2cm,headheight=17pt,headsep=10pt,footskip=15pt]{geometry} 
\usepackage{parskip}
\usepackage{amsmath,amsthm,amssymb,amsfonts}
\usepackage{textcomp}
\usepackage{braket}
\usepackage{bbold}
\usepackage{siunitx}
\usepackage{graphicx}
\usepackage{xcolor}

\newcommand{\bfvec}[1]{\mathbf{#1}}
\newcommand{\bfunitvec}[1]{\hat{\mathbf{#1}}}
\newcommand{\sphericaltensor}[2]{\mathrm{T}^{(#1)}(#2)}
\newcommand{\threejsymbol}[6]{\begin{pmatrix}#1 & #2 & #3 \\ #4 & #5 & #6 \end{pmatrix}}
\newcommand{\sixjsymbol}[6]{\begin{Bmatrix}#1 & #2 & #3 \\ #4 & #5 & #6 \end{Bmatrix}}
\newcommand{\ninejsymbol}[9]{\begin{Bmatrix}#1 & #2 & #3 \\ #4 & #5 & #6 \\ #7 & #8 & #9 \end{Bmatrix}}

\begin{document}

\title{Measuring the nuclear magnetic quadrupole moment in heavy polar molecules}

\author{C. J. Ho}\altaffiliation{Present address: Cavendish Laboratory, University of Cambridge, J. J. Thomson Avenue, Cambridge CB3 0HE, UK}
\author{J. Lim}
\author{B. E. Sauer}
\author{M. R. Tarbutt}

\affiliation{Centre for Cold Matter, Blackett Laboratory, Imperial College London, Prince Consort Road, London SW7 2AZ, UK}
\date{\today}

\begin{abstract}
Theories that extend the Standard Model of particle physics often introduce new interactions that violate charge-parity (CP) symmetry. CP-violating effects within an atomic nucleus can be probed by measuring its nuclear magnetic quadrupole moment (MQM). The sensitivity of such a measurement is enhanced when using a heavy polar molecule containing a nucleus with quadrupole deformation. We determine how the energy levels of a molecule are shifted by the MQM and how those shifts can be measured. The measurement scheme requires molecules in a superposition of magnetic sub-levels that differ by many units of angular momentum.  We develop a generic scheme for preparing these states. Finally, we consider the sensitivity that can be reached, showing that this method can reduce the current uncertainties on several CP-violating parameters.
\end{abstract}

\maketitle

\section{Introduction}

The observed baryon asymmetry in the Universe is one of the important unsolved problems in cosmology. Possible models that can lead to this asymmetry are discussed in Refs.~\cite{Dolgov1981,KolbTurner1990}. One popular class of models was first described by Sakharov, where the asymmetry can arise under conditions of baryon number violation, C and CP-violation, and thermal non-equilibrium~\cite{Sakharov1991}. The Standard Model cannot account for the observed asymmetry~\cite{Canetti2012}, motivating theories beyond the Standard Model to introduce new sources of CP-violation~\cite{Bodeker2021}. Equivalently, since these theories respect CPT symmetry~\cite{Luders1957}, they introduce new sources of T-violation. An elementary particle with a permanent electric dipole moment (EDM) violates both P and T symmetries~\cite{Purcell1950,Landau1957}, and measurements of these EDMs have been exceptionally fruitful in testing physics beyond the Standard Model. Paramagnetic molecules such as $^{174}$YbF~\cite{Hudson2011}, $^{232}$ThO~\cite{Andreev2018} and $^{180}$HfF$^+$~\cite{Cairncross2017} are excellent probes of P,T-violating physics. These molecules are primarily sensitive to the electron EDM, $d_e$, and the scalar P,T-violating electron-nucleon interaction, $C_S$. The most precise measurement thus far, in ThO, constrains new CP-violating interactions to mass scales above 30~TeV in some models and above 3~TeV in most~\cite{Andreev2018}. 

These electron EDM measurements use paramagnetic molecules with a heavy atom of zero nuclear spin, $I=0$. As a result, they are not sensitive to P,T-violating nuclear moments\footnote{The light atom in the molecule often has nuclear spin but its contribution to any P,T-violating signal is heavily suppressed.}. Instead, one can use isotopes with nuclear spin to probe CP-violating effects within the nucleus. The lowest-order P,T-violating nuclear moment is the EDM (which requires $I \geq 1/2$), but that is effectively screened by the outer electrons~\cite{Schiff1963}. The screening is incomplete due to the finite size of the nucleus~\cite{Hinds1997}, but in paramagnetic atoms and molecules the remaining effect (the nuclear Schiff moment) is negligible compared to the contribution from the electron EDM. Conversely, in diamagnetic atoms and molecules the contribution from the electron EDM is small and the Schiff moment can be probed. At present, the most stringent limits on the size of nucleon EDMs and P,T-violating nucleon-nucleon interactions come from a measurement of the nuclear Schiff moment of $^{199}$Hg~\cite{Graner2016}, a diamagnetic atom with $I = 1/2$. An experiment is being constructed that aims to improve on these constraints by measuring the Schiff moment of $^{205}\mathrm{Tl}$ in a beam of TlF molecules, taking advantage of the strong polarisation in an electric field due to the small splitting between opposite parity rotational states~\cite{Grasdijk2021}. Since the Schiff moment is a screened effect, it is interesting to consider the effect of higher-order moments. The next-order nuclear moment of interest is the magnetic quadrupole moment (MQM) (requires $I \geq 1$), which is not screened and can contribute significantly to the P,T-violating energy shifts of atoms and molecules.

A spherical nucleus can acquire an MQM from the EDM of its valence proton or neutron~\cite{Khriplovich1976}, or from P,T-odd nucleon-nucleon interactions involving the valence nucleon~\cite{Sushkov1984}. If we denote the total nuclear MQM as $M$ and the contribution to the MQM from a single proton or neutron as $M_0^{p/n}$, then for these nuclei we have $M \sim M_0^{p/n}$. In nuclei with quadrupole deformation, there are many nucleons in open shells which collectively contribute to the nuclear MQM, which can increase $M$ by an order of magnitude~\cite{Flambaum2014,Lackenby2018}. Furthermore, the interaction of the nuclear MQM with the magnetic moment of the valence electron in the molecule is enhanced by relativistic effects due to the heavy atom~\cite{Sushkov1984}. Table~\ref{tab:MoleculeMQMs} summarizes these effects for isotopologues of molecules currently used for electron EDM experiments. 

\begin{table}[tb]
\centering
 \begin{tabular}{| c | c | c | c | c | c |} 
 \hline
 Molecule & $I^\pi$ & $\beta_2$ (Ref.~\cite{Moller2016}) & $M$ (Ref.~\cite{Flambaum2014}) & $M$ (Ref.~\cite{Lackenby2018}) & $W_M$ ($10^{33} \si{Hz/(e.cm^2)}$)\\
 \hline
$^{137}$BaF & $\frac{3}{2}^+$ & 0.053 & $0M_0^p-1.2M_0^n$ & -- & -0.385~\cite{Denis2020} \\
$^{173}$YbF & $\frac{5}{2}^-$ & 0.300 & $-10M_0^p-10M_0^n$ & $14M_0^p+26M_0^n$ & -1.055~\cite{Denis2020} \\
$^{173}$YbOH & $\frac{5}{2}^-$ & 0.300 & $-10M_0^p-10M_0^n$ & $14M_0^p+26M_0^n$ & -1.067~\cite{Denis2020} \\
$^{229}$ThO & $\frac{5}{2}^+$ & 0.184 & $0M_0^p-19M_0^n$ & $13M_0^p+27M_0^n$ & 1.10~\cite{Skripnikov2014,Skripnikov2015b} \\
$^{177}$HfF$^+$ & $\frac{7}{2}^-$ & 0.277 & $-19M_0^p-14M_0^n$ & $17M_0^p+42M_0^n$ & 0.494~\cite{Skripnikov2017} \\
$^{179}$HfF$^+$ & $\frac{9}{2}^+$ & 0.267 & $-13M_0^p-13M_0^n$ & $20M_0^p+50M_0^n$ & 0.494~\cite{Skripnikov2017} \\
$^{229}$ThF$^+$ & $\frac{5}{2}^+$ & 0.184 & $0M_0^p-19M_0^n$ & $13M_0^p+27M_0^n$ & 0.88~\cite{Skripnikov2015} \\
 \hline
 \end{tabular}
 \caption{Properties of certain molecules relevant for a nuclear MQM measurement. The selected molecules are ones currently being used or explored for electron EDM measurements. The P,T-violating energy shift due to a nuclear MQM, $M$, is proportional to $W_M M$, where $W_M$ is the interaction strength. We list the nuclear spin, parity, quadrupole deformation ($\beta_2$), $M$ and $W_M$.  The collective enhancement of $M$ due to the deformation of the nucleus was calculated in \cite{Flambaum2014} and \cite{Lackenby2018} using different nuclear orbitals, and expressed in terms of the single-nucleon contributions to the MQM, $M_0^{p/n}$.}
 \label{tab:MoleculeMQMs}
\end{table}

In this paper, we consider how to make a measurement of nuclear MQMs using heavy, polar molecules. We first calculate the size of energy shifts in various states of a molecule due to the nuclear MQM. We then describe a procedure for preparing the molecules in the states most sensitive to the MQM. In particular, we introduce a generic state preparation and readout method which utilises the large tensor Stark shifts in molecules to induce a high-order coupling between the two states used for the measurement. Finally, we assess the sensitivity of such an experiment to CP-violating parameters in the hadronic sector, and compare this to other experiments such as Schiff moment or neutron EDM measurements. Our analysis applies to a wide range of molecules, though we will often use $^{173}$YbF as a specific example. The nuclear spin of $^{173}$Yb is 5/2, while that of $^{19}$F is 1/2. We note that the advantages of polyatomic molecules for measuring MQMs and other CP-violating effects are outlined in Ref.~\cite{Hutzler2020}. 

\section{MQM energy shift in molecules}\label{sec:energy_shifts}

We consider a molecule in a $^2\Sigma$ state with electron spin $\bfvec{S}$ and rotational angular momentum $\bfvec{N}$. The nucleus of interest has spin $\bfvec{I}$. The molecule is in an electric field $\bfvec{{\mathcal E}}$ which is parallel to $z$. A suitable effective Hamiltonian is
\begin{equation}
H = H_{\rm rot} + H_{\rm Stark} +H_{\rm hyp}+ H_{\rm M},
\end{equation}
where $H_{\rm rot} = B \bfvec{N}^2$ is the rotational energy, $H_{\rm Stark} = -\bfvec{\mu}\cdot\bfvec{{\mathcal E}}$ describes the Stark shift, $H_{\rm hyp}$ describes the hyperfine interactions, and $H_{\rm M}$ describes the nuclear MQM interaction. Here, $B$ is the rotational constant and $\bfvec{\mu}$ is the electric dipole moment operator.
The nuclear MQM interaction can be written as
\begin{equation}
    H_\mathrm{M} = -\frac{W_M M}{2I(2I-1)}\bfvec{S}\bfunitvec{T}\bfvec{n},
\label{eq:mqmHamiltonian}
\end{equation}
where $\bfvec{n}$ is a unit vector along the internuclear axis and $\bfunitvec{T}$ is a second-rank tensor whose components are $T_{i,j} = I_i I_j + I_j I_i - \frac{2}{3}\delta_{i,j}I(I+1)$~\cite{Sushkov1984,Flambaum2014}. With the help of appendices 8.1 and 8.2 of Ref.~\cite{BrownCarrington2003}, we can re-write this in spherical tensor notation as 
\begin{equation}
    H_\mathrm{M} = \frac{W_M M}{2I(2I-1)}\sqrt{\frac{20}{3}}\sphericaltensor{1}{\bfvec{S},\bfvec{I}^{(2)}}\cdot\bfvec{n} \equiv \bfvec{Q}\cdot\bfvec{n} = Q_{z'},
\label{eq:mqmHamiltonian2}
\end{equation}
%where $\bfvec{I}^{(2)} \equiv \sphericaltensor{2}{\bfvec{I},\bfvec{I}}$ and $\sphericaltensor{k}{\bfvec{A}_1^{(k_1)},\bfvec{A}_2^{(k_2)}}$ denotes a rank-$k$ spherical tensor formed from the product of tensors $\bfvec{A}_1^{(k_1)} \equiv \sphericaltensor{k_1}{\bfvec{A}_1}$ and $\bfvec{A}_2^{(k_2)} \equiv \sphericaltensor{k_2}{\bfvec{A}_2}$ respectively. We omit the rank when it is equal to 1. 
where $\sphericaltensor{1}{\bfvec{S},\bfvec{I}^{(2)}}$ is a rank-1 tensor constructed from $\bfvec{S}$ and $\bfvec{I}^{(2)}$; the latter is a rank-2 tensor constructed from the nuclear spin $\bfvec{I}$. We see that the MQM interaction has been simplified to a vector $\bfvec{Q}$ that acts along the symmetry axis of the molecule, $z'$.

To keep the calculation simple, we first neglect $H_{\rm hyp}$. Its effects are considered later. We find it advantageous to use a basis where the rotational and spin wavefunctions are separated, so we choose the basis set denoted by $\ket{G,M_G;N,M_N}$, where $\bfvec{G}=\bfvec{S}+\bfvec{I}$ is the total spin and $M_G$ and $M_N$ are the projections of $\bfvec{G}$ and $\bfvec{N}$ on the laboratory $z$-axis. $H_{\rm rot} + H_{\rm Stark}$ is diagonal in $G$, $M_G$ and $M_N$, but the electric field mixes states of different $N$. After diagonalizing $H_{\rm rot} + H_{\rm Stark}$ we obtain the eigenfunctions $\ket{G,M_G;\tilde{N},M_N} = \sum_N a_N \ket{G,M_G;N,M_N}$ and corresponding eigenvalues $E_{\tilde{N},M_N}$. Here, $\tilde{N}$ is a label which connects a state adiabatically to its field-free state.

\begin{figure}[tb]
\begin{center}
\includegraphics{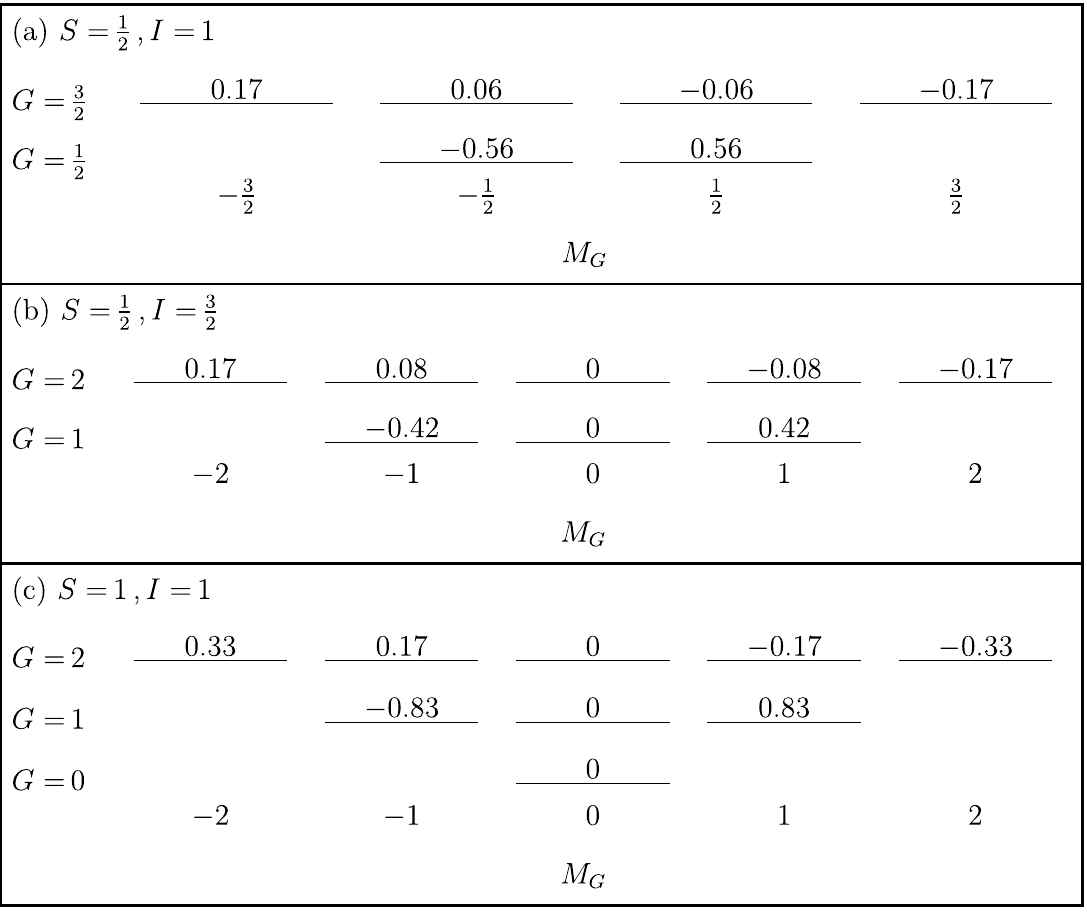}
\caption{Values of $\zeta$ in hyperfine states $\ket{G, M_G}$, for different values of $S$ and $I$. These are calculated using Eq.~(\ref{eq:mqmMESpinPart}).}
\label{fig:mqmShifts}
\end{center}
\end{figure}

The energy shift due to $H_M$ is
\begin{equation}
	\Delta E_\mathrm{M}' = \bra{G,M_G;\tilde{N},M_N}Q_{z'}\ket{G,M_G;\tilde{N},M_N} = W_{\mathrm{M}}M \, \zeta \, \eta.
 \label{eq:dEM1}
\end{equation}
In this equation, $\zeta$ depends only on the electron and nuclear spins and is evaluated in Appendix \ref{sec:app_shifts}, whereas $\eta$ is independent of the spins and expresses the degree of alignment between the internuclear axis of the molecule and the laboratory $z$-axis, often known as the polarization factor. It is given by
\begin{equation}
    \eta = -\frac{1}{\mu_{\rm mol}} \frac{d E_{\tilde{N},M_N}}{d\mathcal{E}},
    \label{eq:eta}
\end{equation}
where $\mu_{\rm mol}$ is the dipole moment along the internuclear axis. Figure~\ref{fig:mqmShifts} shows the evaluation of $\zeta$ for a few selected values of $S$ and $I$. In all the examples given in Fig.~\ref{fig:mqmShifts}, the largest MQM energy shift is found for the states $\ket{G,M_G=\pm G}$, where $G$ does not take its maximum possible value. It is interesting that the configuration where the electron and nuclear spins are parallel is not the one most sensitive to the nuclear MQM. We find this to be true for larger values of $S$ and $I$ as well.

Now we consider the effect of $H_{\rm hyp}$, which includes the spin-rotation interaction, the magnetic dipole hyperfine interaction and the electric quadrupole hyperfine interaction~\cite{Steimle2007}. To handle this, we introduce the spin of the second nucleus $\bfvec{I}_2$, an intermediate angular momentum $\bfvec{F}_1 = \bfvec{G} + \bfvec{N}$ and the total angular momentum $\bfvec{F} = \bfvec{F}_1 + \bfvec{I}_2$. Using the field-free coupled basis, $\ket{N,G,F_1,F,M_F}$, we diagonalize $H_{\rm rot} + H_{\rm Stark} + H_{\rm hyp}$ to find the new eigenstates $\ket{\tilde{N},\tilde{G},\tilde{F}_1,\tilde{F},M_F}$ which are adiabatically linked to the field-free states. Then, we calculate the MQM energy shift 
\begin{equation}
\Delta E_M = \bra{\tilde{N},\tilde{G},\tilde{F}_1,\tilde{F},M_F} H_\mathrm{M} \ket{\tilde{N},\tilde{G},\tilde{F}_1,\tilde{F},M_F}.
\label{eq:mqmEnergyShiftExact}
\end{equation}
The matrix elements needed to calculate $\Delta E_M$ are given in Appendix~\ref{sec:app_shifts}.

Although Eq.~(\ref{eq:mqmEnergyShiftExact}) will be more accurate than Eq.~(\ref{eq:dEM1}), we expect the latter to work well for most $^2\Sigma$ molecules since they have no spin-orbit coupling and the largest spin-rotation coupling is typically much smaller than the hyperfine and rotational energies. As a concrete example, for the $^2\Sigma$ ground state of $^{173}\mathrm{YbF}$, the electron-spin-rotation coupling strength is $\gamma = \SI{-13}{MHz}$, the hyperfine Fermi contact strength between the electron and Yb nuclear spins is $b_F(\mathrm{Yb}) = \SI{-1.98}{GHz}$, and the rotational constant is $B = \SI{7.24}{GHz}$~\cite{Wang2019, Steimle2007}. The fluorine nucleus has spin $I_2 = 1/2$ but this gives rise to a much smaller hyperfine splitting than that of the Yb nucleus ($b_F(\mathrm{F}) = \SI{0.17}{GHz}$). Consequently, the decoupled basis $\ket{N,M_N}\ket{G,M_G}\ket{I_2,M_\mathrm{I_2}}$ is a good first approximation to the exact eigenstates.

\begin{table}[t]
\centering
 \begin{tabular}{| c | c || c | c | c |} 
 \hline
$\ket{\tilde{N},\tilde{G},\tilde{F},M_F}$ & $\Delta E_{\mathrm{M}} (W_M M)$ & $\ket{G,M_G}\ket{I_\mathrm{F},M_{I_\mathrm{F}}}$ & $\zeta$ & $\Delta E_{\mathrm{M}}' (W_M M)$\\
 \hline
$\ket{0,2,5/2,\pm 5/2}$ & $\pm 0.211$ & $\ket{2,\pm 2}\ket{1/2,\pm 1/2}$ & $\pm 14/45$ & $\pm 0.210$ \\
$\ket{0,2,5/2,\pm 3/2}$ & $\pm 0.208$ & $\ket{2,\pm 2}\ket{1/2,\mp 1/2}$ & $\pm 14/45$ & $\pm 0.210$ \\
$\ket{0,2,5/2,\pm 1/2}$ & $\pm 0.094$ & $\ket{2,\pm 1}\ket{1/2,\mp 1/2}$ & $\pm 7/45$ & $\pm 0.105$ \\
$\ket{0,2,3/2,\pm 3/2}$ & $\pm 0.110$ & $\ket{2,\pm 1}\ket{1/2,\pm 1/2}$ & $\pm 7/45$ & $\pm 0.105$ \\
$\ket{0,2,3/2,\pm 1/2}$ & $\pm 0.014$ & $\ket{2,0}\ket{1/2,\pm 1/2}$ & $0$ & $0$ \\ 
& & & & \\
$\ket{0,3,7/2,\pm 7/2}$ & $\mp 0.113$ & $\ket{3,\pm 3}\ket{1/2,\pm 1/2}$ & $\mp 1/6$ & $\mp 0.113$ \\
$\ket{0,3,7/2,\pm 5/2}$ & $\mp 0.079$ & $\ket{3,\pm 2}\ket{1/2,\pm 1/2}$ & $\mp 1/9$ & $\mp 0.075$ \\
$\ket{0,3,7/2,\pm 3/2}$ & $\mp 0.042$ & $\ket{3,\pm 1}\ket{1/2,\pm 1/2}$ & $\mp 1/18$ & $\mp 0.038$ \\
$\ket{0,3,7/2,\pm 1/2}$ & $\mp 0.009$ & $\ket{3,0}\ket{1/2,\pm 1/2}$ & $0$ & $0$ \\
$\ket{0,3,5/2,\pm 5/2}$ & $\mp 0.113$ & $\ket{3,\pm 3}\ket{1/2,\mp 1/2}$ & $\mp 1/6$ & $\mp 0.113$ \\
$\ket{0,3,5/2,\pm 3/2}$ & $\mp 0.076$ & $\ket{3,\pm 2}\ket{1/2,\mp 1/2}$ & $\mp 1/9$ & $\mp 0.075$ \\
$\ket{0,3,5/2,\pm 1/2}$ & $\mp 0.031$ & $\ket{3,\pm 1}\ket{1/2,\mp 1/2}$ & $\mp 1/18$ & $\mp 0.038$ \\
 \hline
 \end{tabular}
 \caption{MQM energy shifts for states correlating to the lowest rotational level of $^{173}\mathrm{YbF}$, in an electric field of $\SI{18}{kV/cm}$. We have dropped $\tilde{F}_1$ from the state notation since $\tilde{F}_1 = \tilde{G}$ for these states. The second column gives $\Delta E_{\mathrm{M}}$ calculated using Eq.~(\ref{eq:mqmEnergyShiftExact}). The fourth column gives $\zeta$ calculated using Eq.~(\ref{eq:mqmMESpinPart}) and the fifth column gives $\Delta E_{\mathrm{M}}'$ calculated using Eq.~(\ref{eq:dEM1}).}
 \label{tab:mqmEnergyShifts}
\end{table}

\begin{figure}[tb]
\begin{center}
\includegraphics{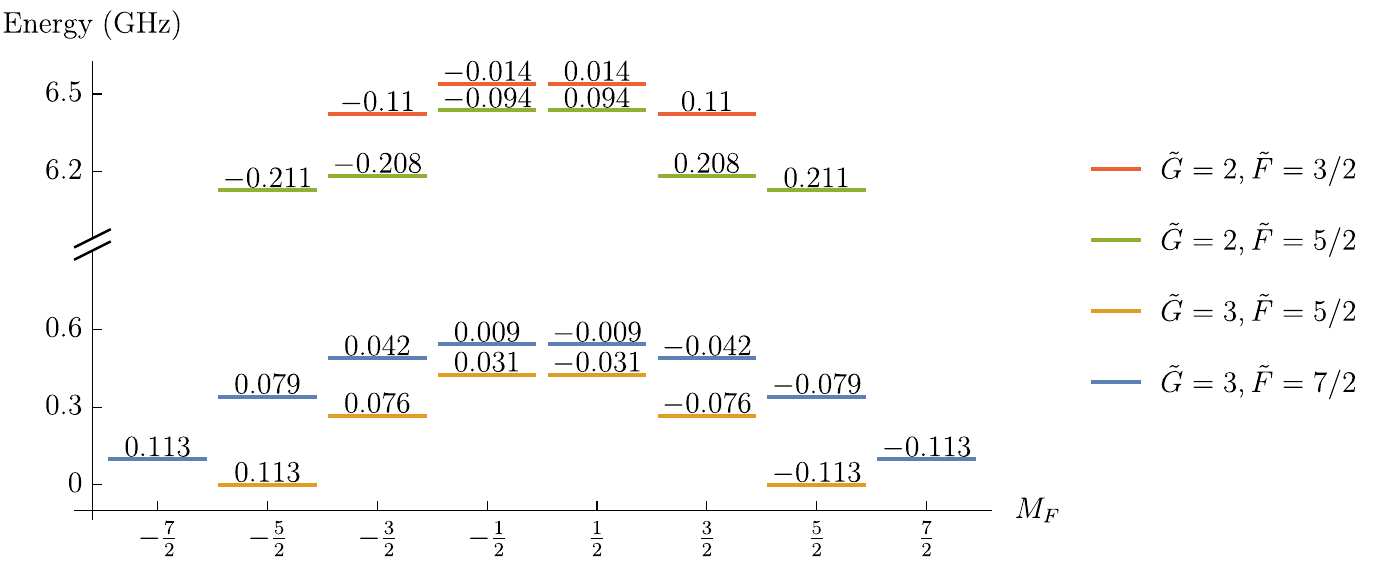}
\caption{MQM energy shifts in the ground rotational state manifold of $^{173}\mathrm{YbF}$. The shifts are calculated at $E=\SI{18}{kV/cm}$ and given in units of $W_M M$. States labeled with $\tilde{G}$ and $\tilde{F}$ are adiabatically linked to field-free states with quantum numbers $G$ and $F$.}
\label{fig:energyLevelsWithMQM}
\end{center}
\end{figure}

Table~\ref{tab:mqmEnergyShifts} lists the MQM energy shifts of all states that correlate to the lowest rotational level of ground state $^{173}\mathrm{YbF}$, in a static electric field of $\SI{18}{kV/cm}$. At this field, the polarization factor is $\eta=0.676$. We have given both $\Delta E_{\mathrm{M}}$ calculated using Eq.~(\ref{eq:mqmEnergyShiftExact}) and the approximate result $\Delta E_{\mathrm{M}}'$ calculated using Eq.~(\ref{eq:dEM1}). The two sets of results are similar, as we would expect from the arguments above. These states, together with the MQM energy shifts, are shown pictorially in Fig.~\ref{fig:energyLevelsWithMQM}. The large splitting between the $\tilde{G}=2$ and $\tilde{G}=3$ states comes from the hyperfine interaction between the electron spin and the Yb nuclear spin. The much smaller splitting into states with $\tilde{F} = \tilde{G} \pm I_2$ is due to the interaction with the F nuclear spin. States of the same $\tilde{F}$ and $\tilde{G}$ but different $|M_F|$ are split by the tensor part of the Stark interaction. This is small for the state with $\tilde{N}=0$, but much larger for all other states. In the absence of the MQM or other $P,T$-violating interactions, states of the same $\{\tilde{G},\tilde{F},|M_F|\}$ are degenerate. The MQM interaction lifts this degeneracy by the amounts indicated in the figure.

\section{Measurement scheme}

An MQM measurement can be done in a similar way to measurements of the electron EDM or the nuclear Schiff moment~\cite{Andreev2018, Ho2020, Grasdijk2021}. Using the notation $\ket{F,M_F}$\footnote{Here, for notational convenience, we have dropped the tildes and the other quantum numbers.}, let us define states $\ket{0} = \ket{F,F}$ and $\ket{\pm} = \frac{1}{\sqrt{2}}\left( \ket{F,+F} \pm \ket{F,-F} \right)$. Molecules are first prepared in $\ket{0}$, typically by optical pumping, and then transferred to $\ket{+}$. This state evolves for time $\tau$ in the presence of parallel electric and magnetic fields, becoming $\ket{\psi} = \frac{1}{\sqrt{2}}\left( e^{i\phi}\ket{F,+F} + e^{-i\phi}\ket{F,-F} \right) = \cos\phi \ket{+} + i \sin\phi\ket{-}$. Here, $\phi = \phi_\mathrm{M} + \phi_\mathrm{B} = \left(\Delta E_\mathrm{M} + \Delta E_\mathrm{B} \right)\tau/\hbar$, where $\Delta E_\mathrm{B}$ is the (absolute) Zeeman shift of the states. Finally, the populations in $\ket{+}$ and $\ket{-}$ are read out, typically by transferring the population in $\ket{+}$ to one state (e.g. back to $\ket{0}$) and the population in $\ket{-}$ to a state of sufficiently different energy that the two are easily resolved (e.g. by laser-induced fluorescence detection).

A particular challenge in this measurement scheme, which does not typically arise in related experiments, is the preparation of a superposition of $\pm M_F$ states where $M_F$ and $-M_F$ differ by several units. For example, consider the levels of $^{173}$YbF shown in Fig.~\ref{fig:energyLevelsWithMQM}. The states most sensitive to the MQM have $\tilde{G}=2,\tilde{F}=5/2,M_F=\pm5/2$. In that case, the measurement scheme calls for a superposition of a pair of states that differ in $M_F$ by 5 units of angular momentum. Previously, optical pumping light modulated at a harmonic of the Larmor frequency has been used to prepare and study coherences within states of high angular momentum~\cite{Yashchuk2003}. In the following, we describe a general method to prepare the states we need.

\subsection{Optical pumping}

We first describe how the molecules can be optically pumped into a single quantum state. This is done at zero electric field. We choose the state $\ket{0}=\ket{F,M_F=F}$, where $F$ is the target angular momentum for the MQM measurement. A small magnetic field defines the $z$-axis. It should be small enough that Zeeman splittings are small compared to the linewidth of the relevant optical transitions, but large enough to ensure that $M_F$ is preserved. Then, using circularly polarized light with $k$-vector along $z$, we drive an optical transition where the excited state angular momentum is $F'=F$. The light drives only $\sigma^+$ transitions so the target state is a dark state and the population will be pumped into this state. There will be decay channels to other hyperfine and rotational states that are not resonant with the optical pumping light. Additional lasers need to be used to drive population out of these states. These extra lasers should have their polarizations modulated at a rate that is close to the Rabi frequency, to ensure that there are no other dark states in the system.

\begin{figure}[tb]
\begin{center}
\includegraphics[width=0.8\textwidth]{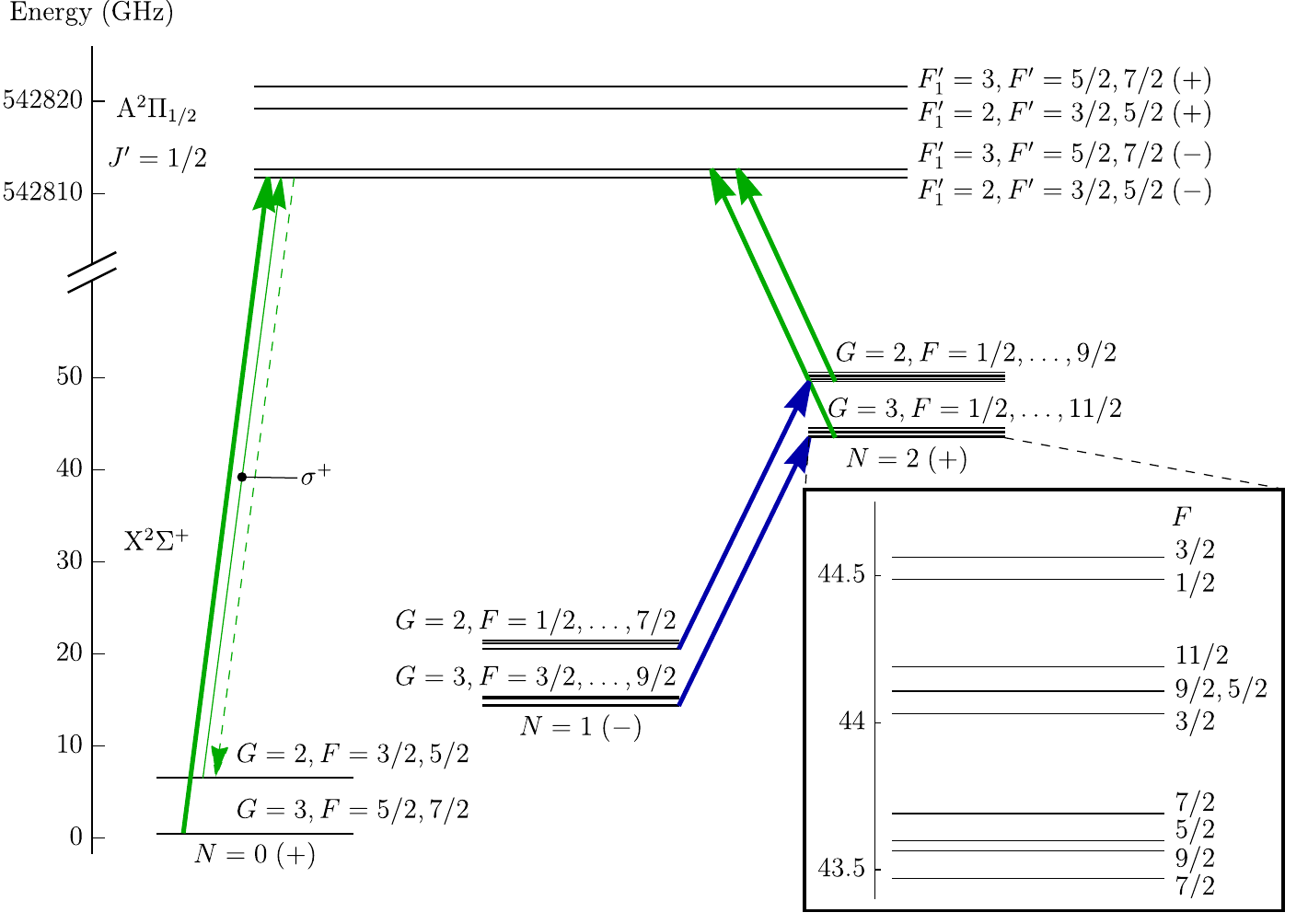}
\caption{Optical pumping scheme for $^{173}$YbF molecules. Thick green arrows denote transitions driven using polarization-modulated light. The thin green arrow denotes light that drives only $\sigma^+$ transitions from the $G=2$ manifold of $N=0$. The target state, $\ket{N=0,G=2,F=5/2,M_F=5/2}$, is the only dark state. Microwave transitions (blue arrows) can also be driven in order to bring population from $N=1$ to the target state. The inset shows the  hyperfine structure within $N=2,G=3$.}
\label{fig:opticalPumping}
\end{center}
\end{figure}

Figure~\ref{fig:opticalPumping} shows an example of this procedure for $^{173}\mathrm{YbF}$ molecules. Within the ground state $\mathrm{X}^2\Sigma^+$, each rotational manifold (labelled by $N$) is split into states $G=2$ and $G=3$ by the Yb hyperfine interaction. Each of these is then further split by spin-rotation and F hyperfine interactions. As an example, the inset shows the hyperfine structure of the $N=2, G=3$ manifold, spanning roughly \SI{1}{GHz}. The figure also shows the lowest rotational manifold of the electronically excited state, $\mathrm{A}^2\Pi_{1/2} (J'=1/2)$. This is split into states of opposite parity by the $\Lambda$-doubling interaction, then by the Yb hyperfine interaction yielding states labelled by $F_1'$, and finally by the fluorine hyperfine interaction, into states of total angular momenta $F'$. The latter splittings are smaller than the linewidth of the optical transition. The target state is $\ket{N=0,G=2,F=5/2,M_F=5/2}$. To optically pump molecules into this state, $\sigma^+$ transitions are driven from the $N=0,G=2$ levels of X to the $F'=3/2,5/2$ levels of A. This state can decay to the $G=3$ manifold of $N=0$ and also to rotational state $N=2$. Molecules decaying to those states are excited by additional polarization-modulated lasers in such a way that the target state is the only dark state. In addition, population in $N=1$ can also be transferred to the target state by driving microwave transitions to $N=2$ as shown. In this way, population distributed across many initial states is all driven to the target state. The efficiency of this process is limited by leaks to higher-lying vibrational states (not shown in the figure). We note that the number of spontaneous emission events needed to reach the stretched state can be substantially reduced by using a combination of coherent and incoherent processes, and that such schemes could improve the optical pumping efficiency in the case of substantial leaks to other states~\cite{Rochester2016}.

\subsection{State preparation and readout}

\begin{figure}[t]
\begin{center}
\includegraphics{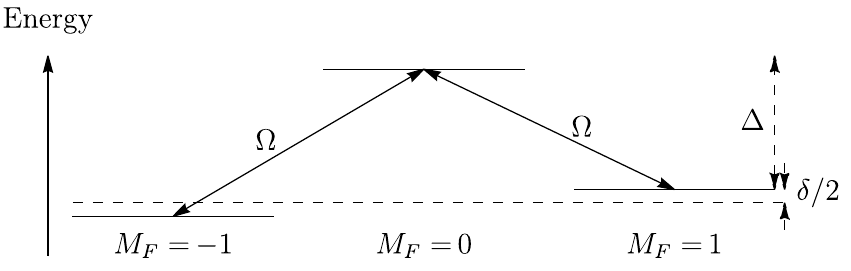}
\caption{An $F=1$ system with tensor Stark splitting $\Delta$ and couplings $\Omega$ between the $M_F=0$ sublevel and $M_F=\pm 1$ sublevels. The $M_F=\pm 1$ states have a splitting $\delta$ between them.}
\label{fig:simpleRaman}
\end{center}
\end{figure}

Next, we show how to drive effective Rabi oscillations between the initial state $\ket{F,M_F= F}$ and the superposition state $\ket{+} = \frac{1}{\sqrt{2}}\left( \ket{F,+F} + \ket{F,-F} \right)$. We start by considering the $F=1$ system illustrated in Fig.~\ref{fig:simpleRaman}. An electric field is applied along $z$ so that there is a Stark shift $\Delta$ between the $M_F=0$ and $|M_F|=1$ states. We also allow a small splitting $\delta$ between the $M_F=\pm1$ states due to a small magnetic field along $z$. There is a time-independent coupling, $\Omega$, between the $M_F=0$ and $M_F = \pm 1$ states. For example, this could be a magnetic field along the $x$-axis, $B_x$, which yields $\Omega = -\mu B_x/\sqrt{2}$, where $\mu$ is the magnetic moment.

The Hamiltonian describing this system, in the basis $\{ M_F=-1, M_F=0, M_F=1 \}$, is
\begin{equation}
	H = \begin{pmatrix}
	-\delta/2 & \Omega & 0 \\
	\Omega & \Delta & \Omega \\
	0 & \Omega & \delta/2
	\end{pmatrix}.
\end{equation}
This is the same Hamiltonian as for a two-photon Raman transition after transforming to the rotating frame. In that situation, $\Omega$ is the Rabi frequency, $\Delta$ is the one-photon detuning and $\delta$ is the two-photon or Raman detuning. Motivated by this, we consider the situation where $\Delta \gg \Omega,\delta$ and the initial state is one of the $M_F=\pm 1$ states. Then, adiabatic elimination of the $M_F=0$ state reduces the dynamics to that of an effective two-level system:
\begin{equation}
	H_\mathrm{eff} = \begin{pmatrix}
	-\frac{\delta}{2} - \frac{\Omega^2}{\Delta} & - \frac{\Omega^2}{\Delta} \\
	- \frac{\Omega^2}{\Delta} & \frac{\delta}{2} - \frac{\Omega^2}{\Delta}
	\end{pmatrix}.
\end{equation}
This system undergoes Rabi oscillations between the $M_F=\pm 1$ states at the effective generalised Rabi frequency, $W_\mathrm{eff} = \sqrt{\delta^2 + \Omega_\mathrm{eff}^2}$, where the effective Rabi frequency is $\Omega_\mathrm{eff} = 2\Omega^2/\Delta$. This effective coupling is mediated by the coupling of $M_F=\pm1$ to $M_F=0$. 

\begin{figure}[t]
\begin{center}
\includegraphics{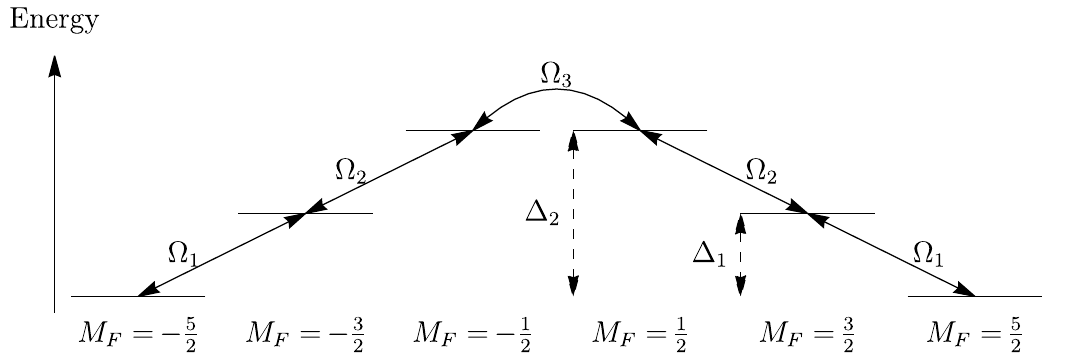}
\caption{An $F=5/2$ system with tensor Stark splittings $\Delta_1$ and $\Delta_2$, and couplings $\Omega_1$, $\Omega_2$, $\Omega_3$ between states of adjacent $M_F$.}
\label{fig:fullRaman}
\end{center}
\end{figure}

This approach can be generalised to states of higher angular momentum. For example, consider the $F=5/2$ system illustrated in Fig.~\ref{fig:fullRaman}. Here we have two different tensor Stark splittings $\Delta_{1,2}$, and three couplings between adjacent levels $\Omega_{1,2,3}$. As before, these couplings can be generated by a magnetic field orthogonal to the electric field. For simplicity, we have made states of equal $M_F$ degenerate. The Hamiltonian for this six-level system is
\begin{equation}
	H = \begin{pmatrix}
	0 & \Omega_1 & 0 & 0 & 0 & 0 \\
	\Omega_1 & \Delta_1 & \Omega_2 & 0 & 0 & 0 \\
	0 & \Omega_2 & \Delta_2 & \Omega_3 & 0 & 0 \\
	0 & 0 & \Omega_3 & \Delta_2 & \Omega_2 & 0 \\
	0 & 0 & 0 & \Omega_2 & \Delta_1 & \Omega_1 \\
	0 & 0 & 0 & 0 & \Omega_1 & 0
	\end{pmatrix}.
\label{eq:hamFullRaman}
\end{equation}
We adiabatically eliminate all states except for $\ket{M_F=\pm 5/2}$ to derive the effective two-level Hamiltonian
\begin{equation}
	H_\mathrm{eff} = \begin{pmatrix}
	-\frac{\Omega_1^2}{\Delta_1} & \frac{\Omega_1^2\Omega_2^2\Omega_3}{\Delta_1^2\Delta_2^2} \\
	\frac{\Omega_1^2\Omega_2^2\Omega_3}{\Delta_1^2\Delta_2^2} & -\frac{\Omega_1^2}{\Delta_1}
	\end{pmatrix}.
\label{eq:hamEffFullRaman}
\end{equation}
The derivation is given in Appendix~\ref{sec:derivationTwoLevel}. We have assumed that the shifts $\Delta_{1,2}$ are much greater than the couplings  $\Omega_{1,2,3}$. The effective Rabi frequency is 
\begin{equation}
\Omega_\mathrm{eff} = \frac{2\Omega_1^2\Omega_2^2\Omega_3}{\Delta_1^2\Delta_2^2}.
\label{eq:Omega5}
\end{equation}

\begin{figure}[t]
\begin{center}
\includegraphics{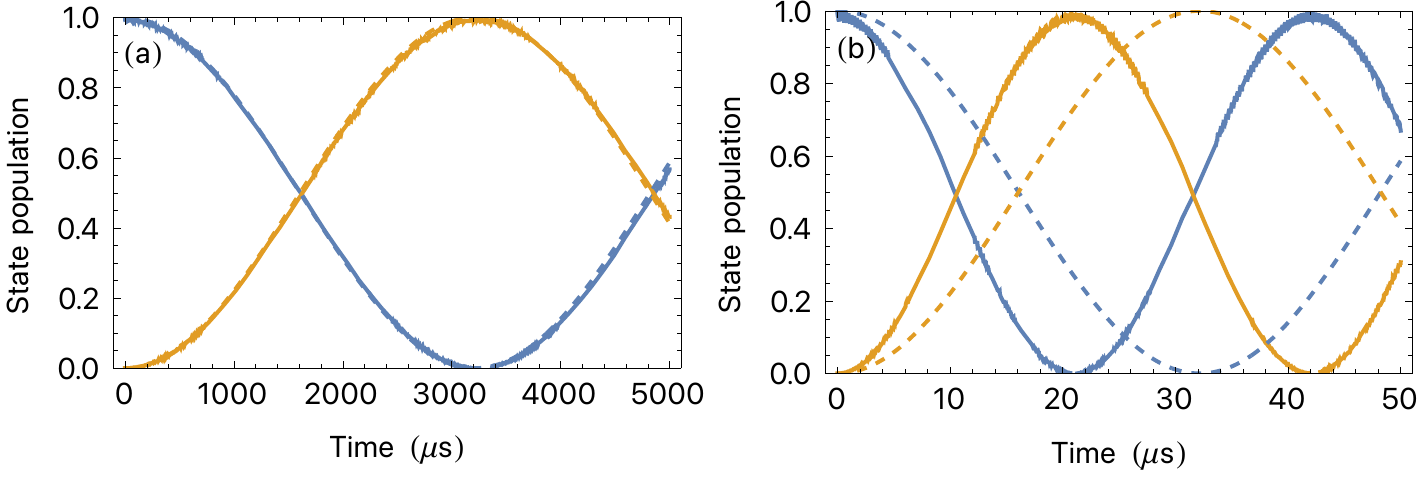}
\caption{Dynamics of the six-level system shown in Fig.~\ref{fig:fullRaman}. Populations of the states $M_F = 5/2$ (blue) and $M_F=-5/2$ (orange). Solid lines: solutions of the Schr\"{o}dinger equation for the Hamiltonian given in Eq.~(\ref{eq:hamFullRaman}). Dashed lines: effective two-level model, Eq.~(\ref{eq:hamEffFullRaman}). Parameters are: (a) $\Omega_1 = \Omega_2 = \Omega_3 = 2\pi\times\SI{6}{MHz}$, $\Delta_1=\Delta_2=2\pi\times\SI{100}{MHz}$. (b) Same as (a) except $\Delta_2 = 2\pi\times\SI{10}{MHz}$.}
\label{fig:fullRamanSims}
\end{center}
\end{figure}

Figure~\ref{fig:fullRamanSims}~(a) compares the numerical solution of the Schr\"{o}dinger equation using Eq.~(\ref{eq:hamFullRaman}) to the effective two-level dynamics described by Eq.~(\ref{eq:hamEffFullRaman}). We see excellent agreement between the effective model and the complete solution. We also find that two-level Rabi flopping dynamics are obtained whenever $\Delta_1 \gg \Omega_1$, without needing to constrain $\Delta_2, \Omega_2, \Omega_3$. The first condition ensures that the amplitudes of all intermediate states remain small, even when $\Omega_2$ and $\Omega_3$ are large. This greatly increases the effective Rabi frequency, producing a rapid coupling between the stretched states even though this is a high-order process mediated through many intermediate states. Note that in this case the effective Rabi frequency is no longer given by Eq.~(\ref{eq:Omega5}). Figure~\ref{fig:fullRamanSims}~(b) compares the numerical solution to the two-level model in the case where the coupling between the states is similar to $\Delta_2$. The numerical solution shows almost perfect Rabi oscillations between the stretched states, with negligible population in any of the other states. The Rabi frequency is larger than given by Eq.~(\ref{eq:Omega5}).

\begin{figure}[t]
\begin{center}
\includegraphics{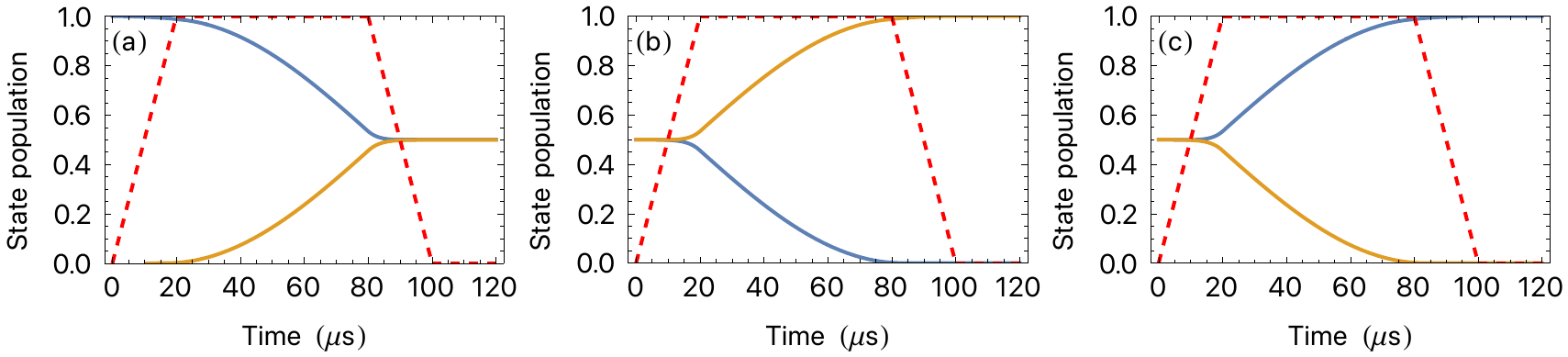}
\caption{Dynamics within the $G=2$ manifold of $^{173}$YbF induced by a pulse of magnetic field orthogonal to the electric field, calculated by numerical solution of the Schr\"odinger equation for all 10 levels shown in the upper half of Fig.~\ref{fig:energyLevelsWithMQM}. Populations of the states $M_F = 5/2$ (blue) and $M_F=-5/2$ (orange) as  $B_x$ is ramped on and off for a total time of $\SI{100}{\mu s}$. The maximum value of the field is $\SI{5.9}{mT}$ and its amplitude relative to this maximum is shown by the dashed red line.
(a) Initial state is $\ket{5/2,5/2}$; final state is $\ket{+}$. (b) Initial state is $\ket{+}$; final state is $\ket{5/2,-5/2}$.  (c) Initial state is $\ket{-}$; final state is $\ket{5/2,5/2}$.}
\label{fig:YbFRaman}
\end{center}
\end{figure}

Let us consider again the specific example of the $G=2$ manifold of $^{173}\mathrm{YbF}$. The upper half of Fig.~\ref{fig:energyLevelsWithMQM} shows the level structure of this manifold of 10 states in an electric field $E=\SI{18}{kV/cm}$. We start with all population in the state $\ket{F,M_F}=\ket{5/2,5/2}$. We apply a magnetic field along $x$, $B_x$, such that $\Omega_{ij} = - \mu_{ij}B_x$ where $\mu_{ij}$ is the magnetic dipole transition moment between states $i$ and $j$. We ramp the field on for $t_\mathrm{ramp} = \SI{20}{\mu s}$, keep the field at its maximum value for $t_\mathrm{on} = \SI{60}{\mu s}$ and then ramp the field down for $t_\mathrm{ramp}$. The total time for the state evolution is therefore $\SI{100}{\mu s}$. Figure~\ref{fig:YbFRaman} shows our numerical simulations for $B_x = \SI{5.9}{mT}$, which is the value needed to apply an effective $\pi/2$ pulse. We find that when the initial state is $\ket{5/2,5/2}$ the final state is $\ket{+}$ as desired. The reverse process also works well, as is needed for the state readout. When the initial state is $\ket{+}$ the final state is $\ket{5/2,-5/2}$, and when the initial state is $\ket{-}$ the final state is $\ket{5/2,5/2}$. Thus, despite the complexity of this system, a magnetic field pulse is sufficient to prepare the initial superposition state and to read out the final state. 

\section{Sensitivity to CP-violating parameters}

Finally, we consider the sensitivity of a molecular MQM experiment to hadronic CP-violating parameters, and compare related experiments. Again, we use  $^{173}\mathrm{YbF}$ as a prototypical example. The parameters of the experiment are the coherence time $\tau$ and the number of molecules measured $N$. At the shot-noise limit, the statistical uncertainty of an ideal frequency measurement will be $\delta f = 1/(2\pi\tau \sqrt{N})$. Recent experimental work and case studies~\cite{Alauze2021,Grasdijk2021,Fitch2021,Augenbraun2020} suggest that measurements using the molecules considered here can reach a sensitivity of approximately $1$~mHz/$\sqrt{\mathrm{Hz}}$. This could be achieved using a high flux focussed molecular beam from a cryogenic source, giving $N \approx 3 \times 10^8$ molecules per second and $\tau \approx 10$~ms, or an ultracold molecular beam giving $N \approx 3 \times 10^6$ molecules per second and $\tau \approx 100$~ms, or ultracold molecules in an optical lattice giving $N \approx 3 \times 10^4$ molecules per second and $\tau \approx 1$~s. In this case, the uncertainty reaches $\delta f = 1$~$\mu$Hz in about 300 hours of measurement time. 

If we assume that the CP-violating energy splitting in $^{173}$YbF is due only to the MQM of the Yb nucleus, $2\Delta_\mathrm{CP} = 0.42 W_M M$, then the uncertainty is equivalent to a $2\sigma$-sensitivity on the MQM of $\SI{4.5e-13}{e.fm^2}$. From this, we can calculate the expected level to which we can measure various CP-violating parameters, using the parameter sensitivities compiled in Table~\ref{tab:sensitivityToCPVParams}. We compare these to the experimental constraints set by the Schiff moment measurement of $^{199}$Hg~\cite{Graner2016} and the projected sensitivities of the proposed Schiff moment measurement of $^{205}$Tl from a TlF beam experiment~\cite{Grasdijk2021}. The former is $S(^{199}\mathrm{Hg}) < \SI{3.1e-13}{e.fm^3}$, and the latter has an anticipated statistical sensitivity of $\delta f = \SI{45}{nHz}$ which corresponds to\footnote{This calculation comes from $\Delta_\mathrm{CP} = \eta W_S S$, where the polarisation factor $\eta = 0.547$ and $W_S = \SI{40539}{a.u.} = \SI{1.8e6}{Hz/(e.fm^3)}$~\cite{Flambaum2020a}.} $\delta S(^{205}\mathrm{Tl}) = \SI{9.1e-14}{e.fm^3}$.

Table~\ref{tab:constraintsCPVParams} shows the 95\% confidence limit constraints on CP-violating parameters set by the $^{199}$Hg experiment, as well as projected $2\sigma$-sensitivities for experiments using TlF and YbF, given the uncertainties quoted above. We find that a measurement of the MQM of $^{173}$Yb using YbF molecules can make large improvements relative to the current limits on CP-violating parameters obtained from the Hg experiment. The measurement also gives greater sensitivities than obtainable from the proposed TlF experiment, even though the projected frequency uncertainty of the TlF experiment is 20 times smaller. 

\begin{table}[t]
\centering
 \begin{tabular}{| c | c | c | c | c | c |} 
 \hline
 CP-violating parameter ($x$) & $\partial S(^{199}\mathrm{Hg})/\partial x$ & $\partial S(^{205}\mathrm{Tl})/\partial x$ & $\partial M_0^p/\partial x$ & $\partial M_0^n/\partial x$ & $\partial M(^{173}\mathrm{Yb})/\partial x$ \\
 \hline
Isoscalar $\pi NN$ coupling ($\bar{g}_0$) & \SI{0.32}{e.fm^3} & \SI{1.77}{e.fm^3} & \SI{-0.08}{e.fm^2} & \SI{-0.09}{e.fm^2} & \SI{-3.4}{e.fm^2} \\
Isovector $\pi NN$ coupling ($\bar{g}_1$) & \SI{-0.10}{e.fm^3} & \SI{-0.05}{e.fm^3} & \SI{0.41}{e.fm^2} & \SI{0.44}{e.fm^2} & \SI{17}{e.fm^2} \\
Isotensor $\pi NN$ coupling ($\bar{g}_2$) & \SI{0.39}{e.fm^3} & \SI{-3.67}{e.fm^3} & \SI{0.16}{e.fm^2} & \SI{0.18}{e.fm^2} & \SI{6.88}{e.fm^2} \\
QCD $\theta$-term ($\tilde{\theta}$) & \SI{0.005}{e.fm^3} & \SI{0.027}{e.fm^3} & \SI{0.002}{e.fm^2} & \SI{0.003}{e.fm^2} & \SI{0.09}{e.fm^2} \\
Proton EDM ($d_p$ (\si{e.fm})) & \SI{0.06}{fm^2} & \SI{0.4}{fm^2} & \SI{0.25}{fm} & 0 & \SI{3.5}{fm} \\
Neutron EDM ($d_n$ (\si{e.fm})) & \SI{0.6}{fm^2} & 0 & 0 & \SI{0.25}{fm} & \SI{6.4}{fm} \\
$u$-quark chromo-EDM ($\tilde{d}_u$ (\si{fm})) & 0 & \SI{9}{e.fm^2} & \SI{12}{e.fm} & \SI{13}{e.fm} & \SI{506}{e.fm} \\
$d$-quark chromo-EDM ($\tilde{d}_d$ (\si{fm})) & \SI{5}{e.fm^2} & \SI{12}{e.fm^2} & \SI{-12}{e.fm} & \SI{-13}{e.fm} & \SI{-506}{e.fm} \\
 \hline
 \end{tabular}
 \caption{The dependence of the Schiff moments of $^{199}$Hg and $^{205}$Tl~\cite{Flambaum2020}, nucleon MQMs and the MQM of $^{173}$Yb~\cite{Lackenby2018} on CP-violating parameters. The dependence of $M$($^{173}$Yb) is calculated from the enhancement of nucleon MQMs as given in Table~\ref{tab:MoleculeMQMs}. Throughout, we take the strong $\pi$-meson nucleon-nucleon interaction constant to be $g=13.6$~\cite{Flambaum2014}.}
 \label{tab:sensitivityToCPVParams}
\end{table}

\begin{table}[t]
\centering
 \begin{tabular}{| c | c | c | c |} 
 \hline
 CP-violating parameter ($x$) & $^{199}$Hg & $^{205}$TlF & $^{173}$YbF \\
 \hline
Schiff moment ($S$) & \SI{3.1e-13}{e.fm^3} & \SI{9.1e-14}{e.fm^3} & -- \\
Magnetic quadrupole moment ($M$) & -- & -- & \SI{4.5e-13}{e.fm^2} \\
Isoscalar $\pi NN$ coupling ($\bar{g}_0$) & \num{1.0e-12} & \num{5.1e-14} & \num{1.3e-13} \\
Isovector $\pi NN$ coupling ($\bar{g}_1$) & \num{3.3e-12} & \num{1.7e-12} & \num{2.6e-14} \\
Isotensor $\pi NN$ coupling ($\bar{g}_2$) & \num{8.1e-13} & \num{2.4e-14} & \num{6.5e-14} \\
QCD $\theta$-term ($\tilde{\theta}$) & \num{6.2e-11} & \num{3.3e-12} & \num{5.0e-12} \\
Proton EDM ($d_p$) & \SI{5.1e-25}{e.cm} & \SI{2.3e-26}{e.cm} & \SI{1.3e-26}{e.cm} \\
Neutron EDM ($d_n$) & \SI{5.1e-26}{e.cm} & -- & \SI{7.0e-27}{e.cm} \\
$u$-quark chromo-EDM ($\tilde{d}_u$) & -- & \SI{1.0e-27}{cm} & \SI{8.9e-29}{cm} \\
$d$-quark chromo-EDM ($\tilde{d}_d$) & \SI{6.2e-27}{cm} & \SI{7.5e-28}{cm} & \SI{8.9e-29}{cm} \\
 \hline
 \end{tabular}
 \caption{Constraints (95\% CL) on CP-violating parameters set experimentally by $^{199}$Hg~\cite{Graner2016}, and projected $2\sigma$-sensitivities from proposals to measure these in $^{205}$TlF~\cite{Grasdijk2021} and $^{173}$YbF (this paper). For $^{199}$Hg, the constraints given here differ from those given in \cite{Graner2016} because we use the coefficients from  \cite{Flambaum2020}.}
 \label{tab:constraintsCPVParams}
\end{table}

\section{Conclusions}

Measurements of nuclear MQMs in isotopologues of heavy, paramagnetic molecules have the potential to probe hadronic CP-violating physics at a level well beyond current limits. We have calculated the molecular energy shifts to be expected, showing which states should be used. We find that measurements will require molecules prepared in a superposition of magnetic sub-levels that differ by many units of angular momentum, and we have described a general method for preparing these states. These experiments would benefit from the recent advances in laser cooling applied to molecules~\cite{Fitch2021b}. The nuclear spin of the heavy atom in the molecule leads to large hyperfine intervals and a more complex hyperfine structure which makes the laser cooling more difficult than for the isotopologues cooled so far. Nevertheless, a recent study~\cite{Kogel2021} shows how cooling of the required molecules can be done with relatively small additions to existing experiments.

 \begin{acknowledgments}
 
This work was supported by funding in part from the
Science and Technology Facilities Council (grants ST/S000011/1 and ST/V00428X/1); the Sloan Foundation (grant G-2019-12505); and the Gordon and Betty Moore Foundation (grant 8864). The opinions expressed in this publication are those of the
author(s) and do not necessarily reflect the views of these funding bodies.

 \end{acknowledgments}
 
\appendix
\section{Further details on the evaluation of the MQM energy shift}\label{sec:app_shifts}

In Sec.~\ref{sec:energy_shifts}, we considered the energy level shifts of a $^{2}\Sigma$ molecule due to the nuclear MQM interaction. Here, we provide more details of the calculation. As described in Sec.~\ref{sec:energy_shifts}, the Hamiltonian is
\begin{equation}
H = H_{\rm rot} + H_{\rm Stark} +H_{\rm hyp}+ H_{\rm M}.
\end{equation}
In the field-free basis, the matrix elements of $H_{\rm rot}$ are
\begin{equation}
\bra{N,M_N}H_{\rm rot}\ket{N',M_N'} = B N(N+1)\delta_{N,N'}\delta_{M_N,M_N'}.
\end{equation}
We can write $H_{\rm Stark} = -\mu_{\rm mol} {\mathcal E} \mathcal{D}_{00}$ where $\mathcal{D}$ is the rotation operator that transforms from the molecule frame to the laboratory frame. The matrix elements of $H_{\rm Stark}$ are
\begin{equation}
    \bra{N,M_N}H_{\rm Stark}\ket{N',M_N'} = - \mu_{\rm mol}{\mathcal E} (-1)^{M_N}\sqrt{(2N+1)(2N'+1)}\threejsymbol{N}{1}{N'}{-M_N}{0}{M_N'}\threejsymbol{N}{1}{N'}{0}{0}{0}.
\end{equation}

Neglecting the hyperfine interaction, the energy shift due to $H_M$ is
\begin{equation}
	\Delta E_\mathrm{M}' = \bra{G,M_G;\tilde{N},M_N}Q_{z'}\ket{G,M_G;\tilde{N},M_N}.
\end{equation}
This factorizes as
\begin{equation}
    \Delta E_\mathrm{M}' = \bra{G,M_G;\tilde{N},M_N}\mathcal{D}_{00}Q_{z}\ket{G,M_G;\tilde{N},M_N} = \bra{G,M_G}Q_{z}\ket{G,M_G}\bra{\tilde{N},M_N}\mathcal{D}_{00}\ket{\tilde{N},M_N} = (W_{\mathrm{M}}M \zeta) \eta.
\end{equation}
The first factor depends only on electron and nuclear spins and evaluates to
\begin{align}
	\bra{G,M_G}Q_z\ket{G,M_G} &= W_{\mathrm{M}} M \zeta =  W_{\mathrm{M}} M (-1)^{G-M_G}\sqrt{\frac{5}{6}}(2G+1)\sqrt{\frac{ S(S+1)(2S+1)(2I+1)(2I+2)(2I+3)}{2I(2I-1)}} \nonumber \\ 
	& \times \threejsymbol{G}{1}{G}{-M_G}{0}{M_G}\ninejsymbol{G}{G}{1}{S}{S}{1}{I}{I}{2}.
\label{eq:mqmMESpinPart}
\end{align}
The second factor is $\eta =\bra{\tilde{N},M_N}\mathcal{D}_{00}\ket{\tilde{N},M_N}$ and is known as the polarization factor. To evaluate $\eta$, we note that
\begin{equation}
	\langle H_\mathrm{Stark} \rangle = \bra{\tilde{N},M_N} H_\mathrm{Stark} \ket{\tilde{N},M_N} = -\mu_\mathrm{mol}\mathcal{E}\bra{\tilde{N},M_N}\mathcal{D}_{00}\ket{\tilde{N},M_N} = -\eta \mu_\mathrm{mol} \mathcal{E},
\end{equation}
and that
\begin{equation}
    \left\langle \frac{d H_\mathrm{Stark}}{d\mathcal{E}} \right\rangle = \left\langle \frac{d H}{d\mathcal{E}} \right\rangle = \frac{d\langle H \rangle}{d \mathcal{E}} = \frac{d E_{\tilde{N},M_N}}{d\mathcal{E}}.
\end{equation}
Thus, we see that
\begin{equation}
    \eta = -\frac{1}{\mu_{\rm mol}} \frac{d E_{\tilde{N},M_N}}{d\mathcal{E}},
\end{equation}
which is straightforward to calculate once the eigenvalues $E_{\tilde{N},M_N}$ have been found.

When we include the hyperfine interaction and the spin of the second nucleus, we use the field-free basis $\ket{N,G,F_1,F,M_F}$ as discussed in Sec.~\ref{sec:energy_shifts}. The matrix elements of $H_{\rm hyp}$ depend on the terms included in the hyperfine interaction and can all be found in \cite{BrownCarrington2003}. The eigenstates of $H_{\rm rot} + H_{\rm Stark} + H_{\rm hyp}$ are denoted $\ket{\tilde{N},\tilde{G},\tilde{F}_1,\tilde{F},M_F}$ and the energy shift due to the MQM is
\begin{equation}
\Delta E_M = \bra{\tilde{N},\tilde{G},\tilde{F}_1,\tilde{F},M_F} H_\mathrm{M} \ket{\tilde{N},\tilde{G},\tilde{F}_1,\tilde{F},M_F}.
\end{equation}
This can be evaluated with the help of the matrix element
\begin{align}
     \bra{N',G',F_1,F,M_F} H_\mathrm{M} \ket{N,G,F_1,F,M_F} &= W_M M (-1)^{F+2F_1+I_2+G}\sqrt{\frac{5}{6}}\nonumber\\ &\times \sqrt{(2F+1)(2F_1+1)(2G+1)(2G'+1)(2N+1)(2N'+1)}\nonumber\\
     &\times \sqrt{\frac{S(S+1)(2S+1)(2I+1)(2I+2)(2I+3)}{2I(2I-1)}} \nonumber \\
     & \times \sixjsymbol{F_1}{F}{I_2}{F}{F_1}{0}\sixjsymbol{G}{N}{F_1}{N'}{G'}{1} \threejsymbol{N'}{1}{N}{0}{0}{0} \ninejsymbol{G'}{G}{1}{S}{S}{1}{I}{I}{2}.
\end{align}

% To evaluate the matrix elements of Eq.~(\ref{eq:mqmHamiltonian}), we rewrite it in spherical tensor form. Here, we found appendices 8.1 and 8.2 of Ref.~\cite{BrownCarrington2003} helpful. We obtain
% \begin{align}
%     \bfvec{S}\bfunitvec{T}\bfvec{n} &= 
%     \begin{pmatrix} S_x & S_y & S_z \end{pmatrix}
%     \begin{pmatrix}
%         2I_x^2 - \frac{2}{3}I(I+1) & I_xI_y + I_yI_x & I_xI_z + I_zI_x \\
%         I_xI_y + I_yI_x & 2I_y^2 - \frac{2}{3}I(I+1) & I_yI_z + I_zI_y \\
%         I_xI_z + I_zI_x & I_yI_z + I_zI_y & 2I_z^2 - \frac{2}{3}I(I+1)
%     \end{pmatrix}
%     \begin{pmatrix} n_x \\ n_y \\ n_z \end{pmatrix} \nonumber \\
%     &= 2\sphericaltensor{2}{\bfvec{I},\bfvec{I}}\cdot\sphericaltensor{2}{\bfvec{S},\bfvec{n}},
% \label{eq:mqmRewrite1}
% \end{align}
% The right hand side of Eq.~(\ref{eq:mqmRewrite1}) can be written
% \begin{equation}
%     2\sphericaltensor{2}{\bfvec{I},\bfvec{I}}\cdot\sphericaltensor{2}{\bfvec{S},\bfvec{n}} = -\sqrt{\frac{20}{3}}\sphericaltensor{1}{\bfvec{S},\bfvec{I}^{(2)}}\cdot\bfvec{n},
% \label{eq:mqmRewrite2}
% \end{equation}
% where $\bfvec{I}^{(2)} \equiv \sphericaltensor{2}{\bfvec{I},\bfvec{I}}$. Substituting this back into Eq.~\ref{eq:mqmHamiltonian}, we find

\section{Derivation of effective two-level Hamiltonian for $F=5/2$ system}\label{sec:derivationTwoLevel}

The time-dependent Schr\"{o}dinger equation for the Hamiltonian given in Eq.~(\ref{eq:hamFullRaman}) can be written as (setting $\hbar = 1$):
\begin{align*}
	i\dot{a}(t) &= \Omega_1 b(t), \\
	i\dot{b}(t) &= \Omega_1 a(t) + \Delta_1 b(t) + \Omega_2 c(t), \\
	i\dot{c}(t) &= \Omega_2 b(t) + \Delta_2 c(t) + \Omega_3 d(t), \\
	i\dot{d}(t) &= \Omega_3 c(t) + \Delta_2 d(t) + \Omega_2 e(t), \\
	i\dot{e}(t) &= \Omega_2 d(t) + \Delta_1 e(t) + \Omega_1 f(t), \\
	i\dot{f}(t) &= \Omega_1 e(t),
\end{align*}
where the amplitudes $\{a(t), b(t), c(t), d(t), e(t), f(t)\}$ correspond to those of $M_F = -5/2, -3/2, -1/2, 1/2, 3/2, 5/2$ respectively. Consider the initial condition where $f(0) = 1$ and all other amplitudes are zero at $t=0$, and the assumption that $\Delta_1, \Delta_2 \gg \Omega_1, \Omega_2, \Omega_3$. Then, we can approximate $\dot{b}(t) \approx \dot{c}(t) \approx \dot{d}(t) \approx \dot{e}(t) \approx 0$, which gives us
\begin{align*}
	b(t) = -\frac{\Omega_1 a(t) + \Omega_2 c(t)}{\Delta_1}, \\
	c(t) = -\frac{\Omega_2 b(t) + \Omega_3 d(t)}{\Delta_2}, \\
	d(t) = -\frac{\Omega_3 c(t) + \Omega_2 e(t)}{\Delta_2}, \\
	e(t) = -\frac{\Omega_2 d(t) + \Omega_1 f(t)}{\Delta_1}.
\end{align*}
We now substitute the expression for $e(t)$ into $d(t)$ to get
\begin{equation*}
	\left( 1 - \frac{\Omega_2^2}{\Delta_1\Delta_2} \right) d(t) = -\frac{\Omega_3}{\Delta_2}c(t) + \frac{\Omega_1\Omega_2}{\Delta_1\Delta_2}f(t).
\end{equation*}
Since the tensor shifts are much greater than the direct couplings between $M_F$ states, the LHS is approximately $d(t)$. Next, we substitute this expression into the equation for $c(t)$ above and repeat the process. Eventually, we substitute an expression for $b(t)$ into the first differential equation involving $a(t)$ to get
\begin{equation*}
	i\dot{a}(t) = -\frac{\Omega_1^2}{\Delta_1}a(t) + \frac{\Omega_1^2\Omega_2^2\Omega_3}{\Delta_1^2\Delta_2^2}f(t).
\end{equation*}
Similarly, we find for $f(t)$,
\begin{equation*}
	i\dot{f}(t) = \frac{\Omega_1^2\Omega_2^2\Omega_3}{\Delta_1^2\Delta_2^2}a(t) -\frac{\Omega_1^2}{\Delta_1}f(t),
\end{equation*}
which together give the effective two-level Hamiltonian in Eq.~(\ref{eq:hamEffFullRaman}).

\bibliography{references_extra}

\end{document}